# Performance Comparison of Various STM Concurrency Control Protocols Using Synchrobench

Ajay Singh, Sathya Peri, G Monika and Anila kumari
Dept. *of Computer Science And Engineering*
IIT Hyderabad
Hyderabad, India
Emails : {cs15mtech01001, sathya_p, *cs11b044, cs11b043*}@iith.ac.in

*Abstract*— Writing concurrent programs for shared memory multiprocessor systems is a nightmare. This hinders users to exploit the full potential of multiprocessors. STM (Software Transactional Memory) is a promising concurrent programming paradigm which addresses woes of programming for multiprocessor systems.

In this paper, we implement BTO (Basic Timestamp Ordering), SGT (Serialization Graph Testing) and MVTO(Multi-Version Time-Stamp Ordering) concurrency control protocols and build an STM(Software Transactional Memory) library to evaluate the performance of these protocols. The deferred write approach is followed to implement the STM. A *SET* data structure is implemented using the transactions of our STM library. And this transactional *SET* is used as a test application to evaluate the STM. The performance of the protocols is rigorously compared against the linked-list module of the Synchrobench benchmark. Linked list module implements *SET* data structure using lazy-list, lock-free list, lock-coupling list and ESTM (Elastic Software Transactional Memory).

Our analysis shows that for a number of threads greater than 60 and update rate 70%, BTO takes (17% to 29%) and (6% to 24%) less CPU time per thread when compared against lazy-list and lock-coupling list respectively. MVTO takes (13% to 24%) and (3% to 24%) less CPU time per thread when compared against lazy-list and lock-coupling list respectively. BTO and MVTO have similar per thread CPU time. BTO and MVTO outperform SGT by 9% to 36%.

## I. Introduction

With the advent of multi-core processors and in pursuit of harnessing their full potential we need parallel programs, but parallel programming is very challenging in terms of scalability, composability, debugging and of course the difficulty in synchronization of shared memory access. Various problems like deadlocks, priority inversion, convoy effect and data races[15] keep programmers busy in fixing these troubles instead of focusing on logic for parallelization of the applications. All these developments give rise to the need for new parallel programming paradigm.

Software transactional memory (STM) first proposed in 1995, by Nir Shavit and Touitou[16] is promising parallel programming model which makes parallel programming easier and efficient. STM works on concepts of transactions, first proposed for databases. Programmer with help of STMs just needs to identify critical sections within their applications which could be executed as a transaction and all the concurrency issues and dirty work of ensuring correctness, scalability and composition of locks are handled within the STM library, easing the burden on parallel programmers.

In this paper, we compare the Basic Timestamp Ordering (BTO), Serialization Graph Testing (SGT) and Multi-Version Timestamp Ordering (MVTO)[13] concurrency techniques implemented as an STM against the various synchronization techniques of Synchrobench benchmark[12].

Our contribution:
(1) Transactional implementation of the set using linked list with add, delete and contains methods.
(2) Implementation of read/writes model based software transactional memory using state of art concurrency control protocols; BTO, SGT and MVTO[13].
(3) Detailed Performance comparison of BTO, SGT and MVTO of the STM middleware against the set implementation using lazy-list, lock-coupling list, lock-free list and ESTM concurrency control mechanisms of synchrobench[12] benchmark.

The rest of the paper is organized as follows: In section II we briefly discuss the design of our STM middleware and its underlying protocols. In section III we explain the set data structure implementation, which is the test application implemented to test the performance of the STM middleware. In section IV we present the detailed performance of proposed STM implementation against Synchrobench benchmark having lazy-list, lock-coupling, lock-free and ESTM based implementation of a set data structure. The set exposes add, delete and contains methods. In section V we discuss the related literature and finally we conclude with proposal of the future work in section VI.

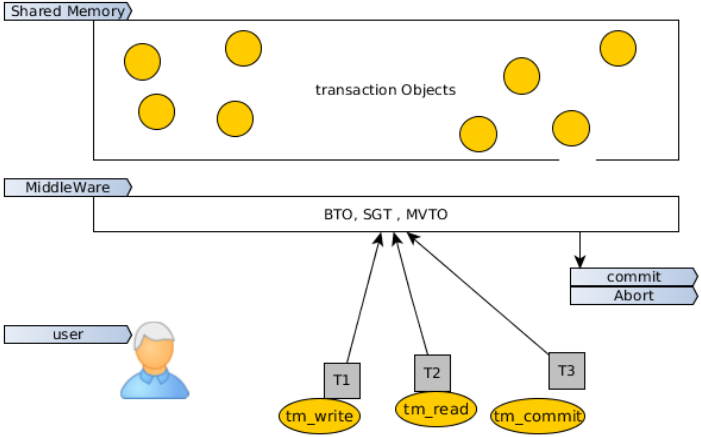

Fig. 1. STM design diagram.

## II. STM Design

The STM system we developed exports *tm_begin*, *tm_read*, *tm_write* and *tm_commit* functions for the parallel programmer. The shared transaction objects or data items reside in shared memory as shown in Figure 1. Each transaction use these exported functions at the user level to access the shared objects. The concurrency issues with concurrent transactions which access the shared objects are handled by the middleware implemented using state of art concurrency control protocols namely- BTO, SGT and MVTO [13]. Shared memory is implemented as thread safe map provided by Intel TBB library [14].

We follow deferred write approach, thus each transaction maintains a local buffer of accessed shared objects during its lifetime. Each read/write request from/to shared objects is locally buffered and updates of all operations are logged locally until the *tm_commit* request for the transaction arrives at middleware. During execution of *tm_commit*, the local log is validated as per the concurrency control protocol used. Once *tm_commit* completes execution and it returns success, we write all the changes to the shared memory atomically. Otherwise, we flush the local log of the transaction and the tm_commit returns an abort.

The Following section requires the use of these definitions and notations:

- **Transaction[13]:** It is shared code segment that needs to execute atomically. And is a finite set of read/write operations on a shared memory.
- **History or Schedule:** A history is interleaving of operations from different transactions. A history is said to be serializable[13] if all the transactions within a history can be serially ordered.
- **Conflicting operations:** Two operations of a history are said to be conflicting, if they access same memory and at least one of them is a write operation.

$T_i$ denotes the $i_{th}$ transaction, $TS(T_i)$ denotes the time-stamp value of $T_i$. Shared objects reside in shared memory. Each transaction operates on shared objects via *tm_begin*, *tm_read*, *tm_write*, *tm_commit* operations. We depict each operation by $p_i(X)$ where $p$ is the shared object on which operations of $T_i$ operate.

In the following subsections we will briefly describe the concurrency control protocols that we implemented as part of the middleware layer as in Figure 1. These protocols are popular in databases [13].

### A. BTO

This protocol follows the time ordering notion according to which if $T_i < T_j$ then all conflicting operation of $T_i$ and $T_j$ follow the transactional order. Thus equivalent serial schedule consists of all operations of $T_i$ ordered before all operations of $T_j$ [13]. Each shared object maintains the timestamp of the last transaction that accessed it for each type of the operation as follows:

- *MAX_READ(X)*: Timestamp of the last transaction that performed *tm_read* on data object *X*.
- *MAX_WRITE(X)*: Timestamp of the last transaction that performed *tm_write* on data object *X*.

For each operation $p_i(X)$ of $T_i$ following are the validation rules of the protocol:

(1) If $p$ is *tm_read*: If $TS(T_i) < MAX\_WRITE(X)$ abort $T_i$ else the operation succeeds.
(2) If $p$ is *tm_write*: log operation locally and validate it during *tm_commit*.
(3) Commit validation: For each data object in the local log, if $TS(T_i) < MAX\_WRITE(X)$ and if $TS(T_i) < MAX\_READ(X)$ abort $T_i$ else $T_i$ commits successfully.

### B. SGT

This protocol maintains a conflict graph, *CG(V, E)* (with *V* as transactions and *E* as conflict edges[13]) and produces an equivalent serial schedule by ensuring that conflict graph is acyclic[13].

Whenever a new operation $p_i(X)$ arrives following steps are taken:

(1) If $p_i(X)$ is the first operation of $T_i$, we add a node in *CG(V, E)*. And add real-time edges from all committed transactions to the vertex represented by $T_i$.
(2) If $p$ is *tm_read*, validation is done as in step 3. If $p$ is *tm_write*, locally log this write and validate during *tm_commit*.
(3) Validation: Assume $q_j(X)$ is the earlier operation and current operation is $p_i(X)$, then an edge $(T_j, T_i)$ is added to the *CG*.
   Now, if the resultant graph is:
   a) Cyclic: $T_i$ is aborted as no serial schedule is possible. Delete node $T_i$ along with all incident edges.
   b) Acyclic: The operation succeeds and the resultant graph is updated as new conflict graph.

To ensure atomic access of the conflict graph mutex locks are used. The conflict graph grows bigger in size as new transactions arrive and commit. This calls for a garbage collection scheme to be implemented so that unnecessary nodes from the graph can be removed. We remove a node $T_i$ from the graph once all the transactions which were active during commit of $T_i$ finish their execution.

*C. MVTO*

Multiversion time-stamp protocol [18] maintains multiple versions of same data objects. We maintain a local read/write log of data items and all writes take effect during *tm_commit*. The written versions of data objects are only available to other transactions after *tm_commit* and each version carries the timestamp of its own transaction. In a case of multiple writes within a transaction, only last value is considered to take effect inside *tm_commit*. In a case of multiple reads, first read operation value is locally logged and subsequent reads use this local value.

Following are the major steps of the implemented protocol for any operation $p_i(X)$:

(1) If $p_i(X)$ is *tm_read*: If the value of $X$ is in the local log, $p_i(X)$ directly return this value. Else $p_i(X)$ reads a value $X_k$ ($K_{th}$ version of $X$) created by $T_k$ such that, $TS(X_k)$ is the largest timestamp $< TS(T_i)$. And $X_k$ is also locally logged.

(2) If $p_i(X)$ is *tm_write*: It creates a new version of $X$ and its value is locally logged.

(3) *tm_commit* of $T_i$: For each $X_k$ written in the local buffer we validate as follows.
   a) If $r_j(X_k)$ has already been scheduled such that $TS(T_k) < TS(T_i) < TS(T_j)$; this implies the version created by $T_i$ is obsolete and it needs to be aborted.
   b) Otherwise, add version $X_i$ to shared memory and it is made visible to other transactions.

```
Algorithm 1 set_add algorithm
1: procedure SET_ADD(val)
2:     T ← begin()                              ▷ transaction begin
3:     set_obj_p ← new common_tOB ▷ initialize prev and next pointers of list
4:     tm_read(T, set_obj_p)
5:     set_obj_n ← next(set_obj_p)
6:     tm_read(T, set_obj_n)
7:     while value(set_obj_n) ≺ val do
8:         set_obj_p ← set_obj_n
9:         set_obj_n ← next(set_obj_n)
10:        tm_read(T, set_obj_n)
11:    if value(set_obj_n)! = val then
12:        set_obj ← newshared_obj(val)
13:        next(set_obj_p) ← set_obj
14:        next(set_obj) ← set_obj_n
15:        tm_write(T, set_obj)
16:        tm_write(T, set_obj_p)
17:        tm_write(T, set_obj_n)
18:    tm_commit(T, error_id)        ▷ commit transaction if write succeeds
```

Fig. 2. Algorithm to add into a set

```
Algorithm 2 set_remove
1: procedure SET_REMOVE(val)
2:     T ← begin()                              ▷ transaction begin
3:     set_obj_p ← new common_tOB ▷ initialize prev and next pointers of list
4:     tm_read(T, set_obj_p)
5:     set_obj_n ← next(set_obj_p)
6:     tm_read(T, set_obj_n)
7:     while true do
8:         v = value(next(set_obj_p))
9:         if v >= val then
10:            break
11:        set_obj_p ← set_obj_n
12:        set_obj_n ← next(set_obj_n)
13:        tm_read(T, set_obj_n)
14:    if v == val then
15:        next(set_obj_p) ← next(set_obj_n)
16:        tm_write(T, set_obj_p)
17:        tm_write(T, set_obj_n)
18:    tm_commit(T, error_id)        ▷ commit transaction if write succeeds
```

Fig. 3. Algorithm to delete from a set.

### III. TEST APPLICATION

We have implemented a *SET* data-structure to test the protocols (BTO, SGT and MVTO) of STM middleware. The *SET* exports *set_add*, *set_contains* and *set_remove* methods. These methods are transactionally implemented as shown in Figure 2, 3 and 4. *SET* is implemented as a linked list. The linked list always is sorted in order of their node values, a node may have values as simple as an integer and as complex as a structure. Without loss of generality and for ease of implementation, we can assume node values to be integers. Each node of the *SET* is a shared object residing in shared memory. The concurrent access to the shared memory has to be synchronized using the middleware protocols. To interact with the *SET* user invokes its exposed methods. Each method is a transaction in itself implemented using the functions exposed by the STM system as shown in Figure 2, 3 and 4. Accessing the *SET* using these transactions ensures that synchronization is handled by the middleware protocols of the STM. Thus, we have built a concurrent *SET* using our STM system.

The *SET* is implemented as a linked list and the shared

```
Algorithm 3 set_contains
1: procedure SET_CONTAINS(val)
2:     T ← begin()                              ▷ transaction begin
3:     set_obj_p ← new common_tOB
4:     tm_read(T, set_obj_p)
5:     set_obj_n ← next(set_obj_p)
6:     tm_read(T, set_obj_n)
7:     while true do
8:         v = value(next(set_obj_p))
9:         if v >= val then
10:            break
11:        set_obj_p ← set_obj_n
12:        set_obj_n ← next(set_obj_n)
13:        tm_read(T, set_obj_n)
14:    if v == val then
15:        Found
16:    else
17:        NotFound
18:    tm_commit(T, error_id)                    ▷ commit transaction
```

Fig. 4. Algorithm to lookup from a set.

objects or common objects are nodes of the linked list. To add, delete or search a node in *SET* with given value we need to traverse the list up to the correct location. Linked list has a structural invariant that it always remains sorted in increasing order of their values. The set initially contains two sentinel nodes with a minimum and maximum value from the range of possible values of the set nodes. Thus we can say shared memory initially contains a *SET* with two sentinel nodes.

*set_add* as shown in Figure 2 adds a value to the set transactionally. Line 2 begins the transaction T and line 3 to 6 read the linked list head and next of head in the local transaction object. *set_obj_p* contains the head of the list and *set_obj_n* points to next node of the list. Line 7 to 10 traverses down to the correct location where a new node with value *val* needs to be added. Note that the nodes are read from the shared memory via *tm_read* method exposed by STM. Line 11 to 14 check if the node (shared object) to be added is already present. If not, it adds the shared object into the local log using *tm_read*. Finally, *tm_commit* writes the local values to the shared memory after the validation. Note that multiple transactions might be executing *set_add* where concurrency issues are handled by the STM library at middleware level. *set_remove* is much similar to *set_add*, as it also traverses the set nodes in the *shared memory* using *tm_read* and *tm_write* methods of STM, locally logging all changes and finally deleting the desired node with value *val* from *shared memory (*underlying set*)* after validation in *tm_commit*. Thus, we see programmer easily focuses on parallelizing the application instead of worrying about the concurrency issues. All concurrency issues are handled inside STM. The Programmer only needs to identify the shared objects and access them within the transactions using *tm_begin, tm_read, tm_write*, and *tm_commit* methods of the STM. *Set_contain* is also much similar to *set_add* and *set_remove* as in Figure 4.

IV. EVALUATION OF STM PROTOCOLS

This section presents a performance comparison of STM middleware protocols namely BTO, MVTO and SGT against lazy-list, ESTM, lock-free list and lock-coupling list respectively. To compare the performance we use the test application discussed in section III. We implement a *SET* using the STM library. And then compare our *SET* implementation against the *SET* implementation of Synchrobench [12] using the lazy-list, ESTM, lockfree list and lock coupling list.

The measurements were taken on Intel(R) Core(TM) i3 CPU with 2 cores, 3.20GHz and 3GB main memory. The system uses Ubuntu version 16.04 for 64bit systems, Glibc version 2.23 and g++ 5.4.0-6. In Figure 9 we measure the execution time of the test application for 100 threads for each protocol using following clock measures: 1) real time taken by application, 2) CPU time taken by the application and 3) per thread CPU time taken. We further present the detailed comparison only for per thread CPU time taken (due to space restrictions) in Figure 5, 6, 7 and 8. Average time (in milliseconds as X-axis) is plotted for 10, 20, 30 and up to 100 threads(as Y axis) respectively. To make the evaluation rigorous, update operation rate is taken 70%. ESTM [17] is evaluated using normal transactions. To measure real time, CPU time taken and per thread time taken by the application, we use CLOCK_MONOTONIC_RAW, CLOCK_PROCESS _CPUTIME_ID and CLOCK_THREAD_CPUTIME_ID clocks of Linux kernel respectively.

We now present STM protocol wise evaluations for 100 threads, please refer Figure 9:

(1) When the measure is real execution time taken by the application, BTO takes 97% and 197% less time than lazy linked list and lock-coupling list(both using spin locks) respectively. On another hand lazy list and lock coupling list using mutex perform similar to BTO, but ESTM and lock-free fare better than BTO. SGT outperforms lazy list and lock-coupling list both using spinlocks by 109% and 176% respectively. And all the other synchrobench *SET* implementations perform badly. Whereas MVTO outperforms lock-coupling list by 130%, and MVTO performs poorly against other synchrobench SET implementations.

(2) When the measure is CPU execution time taken by the application, BTO takes 147% and 198% less time than lazy linked list and lock-coupling list(both using spin locks) respectively. On other hand lazy-list and lock-coupling list via mutex, ESTM and lock-free fare better than BTO. SGT outperforms lazy-list and lock-coupling list both using spinlocks by 58% and 195% respectively. And all other synchrobench *SET* implementations perform badly. Whereas, MVTO outperforms lock-coupling list (using spinlocks) by 90%, and MVTO perform poorly against other synchrobench SET implementations. Please note that in the plots shown in Figure 9 the value of time taken by spin lock coupling and MVTO is more than 450 milliseconds and 1000 milliseconds respectively. These values are scaled down so as to fit in the plot as the purpose is to show relative performance with other protocols and not their absolute performance.

(3) When the measure is per thread CPU execution time, BTO takes 24% and 33% less time than lazy-linked list and lock-coupling list(both using spin locks) respectively. BTO also outperforms other synchrobench protocols using linked list marginally except ESTM which is slightly better than BTO. SGT takes 12% and 2.4% more time for lazy-list and lock-coupling list both using spinlocks respectively. SGT follows same bad performance trend for all other *SET* implementations of synchrobench. Whereas, MVTO outperforms lock-coupling list (via spin-locks), lazy link list (via spinlocks), lock-coupling list (via mutex locks), lazy list (via mutex locks) lock-free list and SGT by 33%, 24%, 22%, 23%, 12% and 36% respectively. ESTM again does better than MVTO.

Amongst BTO, SGT and MVTO - BTO performs better and MVTO fares poorly for all- CPU time taken, the real time taken and per thread CPU time taken measures. The only exception is MVTO, where per thread CPU execution time is 36% better than SGT as in Figure 10. The low performance of SGT and MVTO against BTO might be attributed to the extra overheads of garbage collection in SGT and MVTO, plus the maintenance overheads of extra versions in MVTO. Detailed evaluation of STM middleware protocols against linked list module of Synchrobench can be seen from Figure 5, 6, 7 and 8. We do this only for per thread CPU time measure due to space limit. Source code and detailed documentation of the STM can found at our lab's website [19].

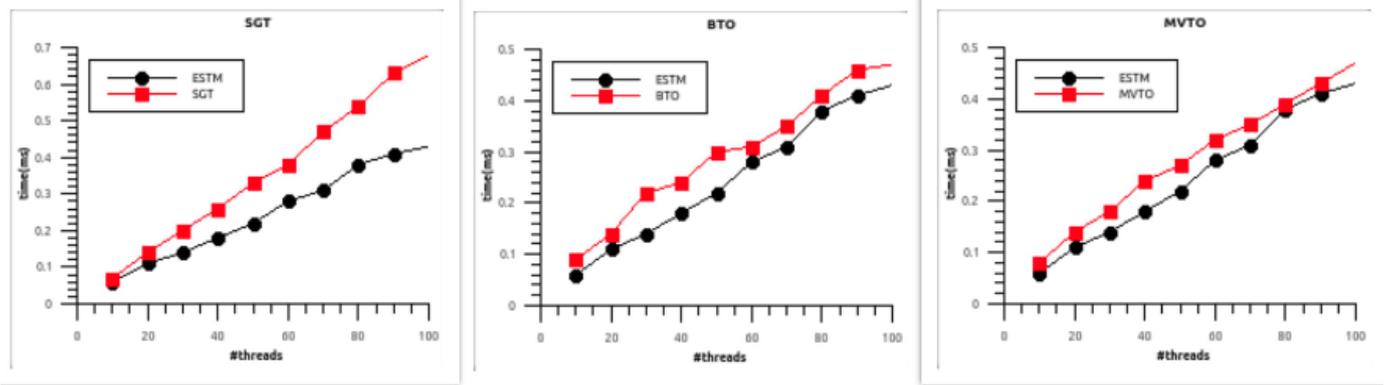

Fig. 5. ESTM Vs {SGT, BTO and MVTO}(left to right). X: number of threads; Y: time in millisecond.

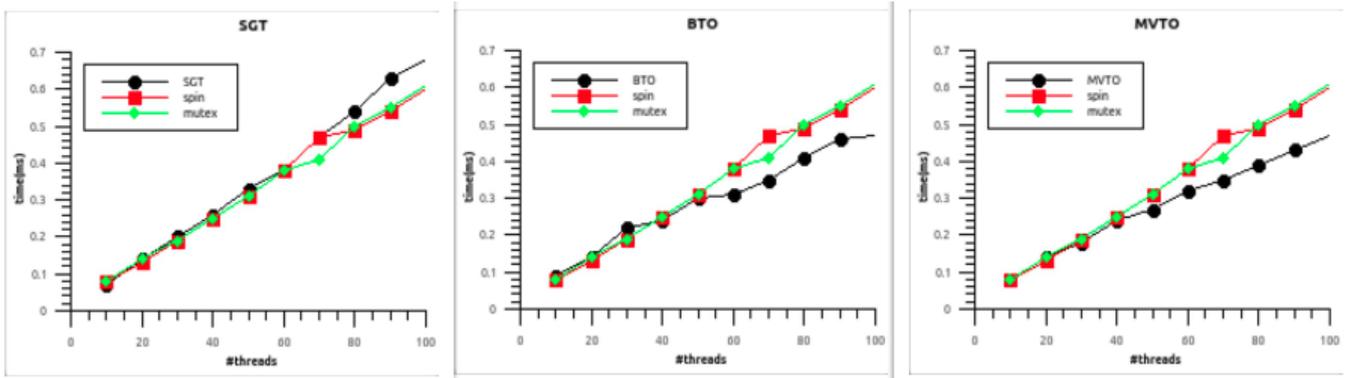

Fig. 6. (Spin-Lazylist, Mutex-Lazylist) Vs (SGT, BTO and MVTO)(left to right). X: number of threads; Y: time in millisecond.

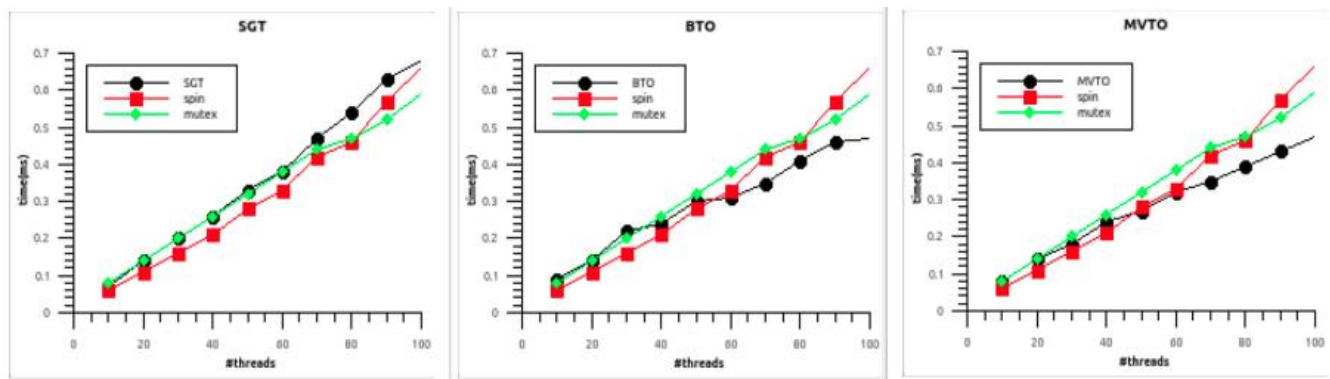

Fig. 7. (Spin-lockcoupling, Mutex-lockcoupling) Vs (SGT, BTO and MVTO)(left to right). X: number of threads; Y: time in millisecond.

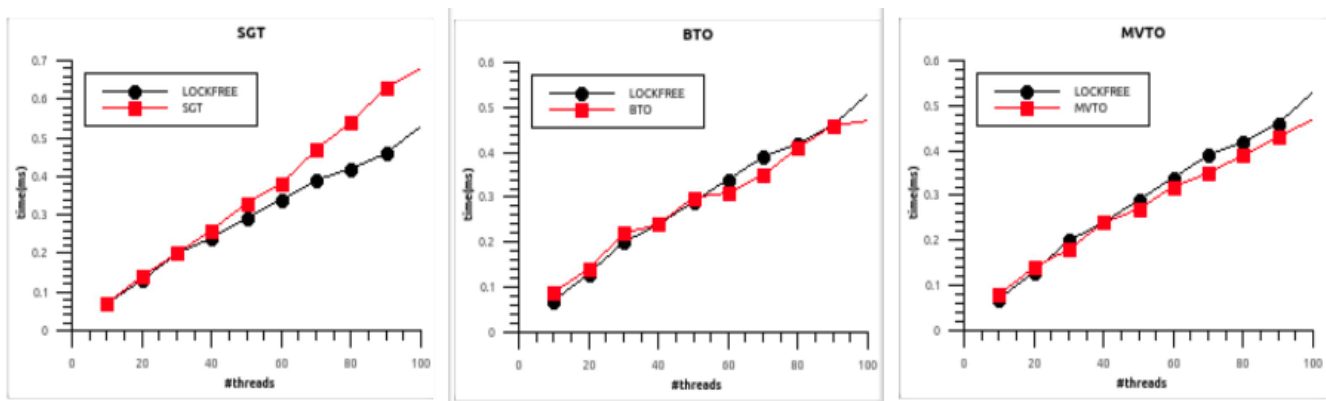

Fig. 8. LOCKFREE Vs {SGT, BTO and MVTO}(left to right). X: number of threads; Y: time in millisecond.

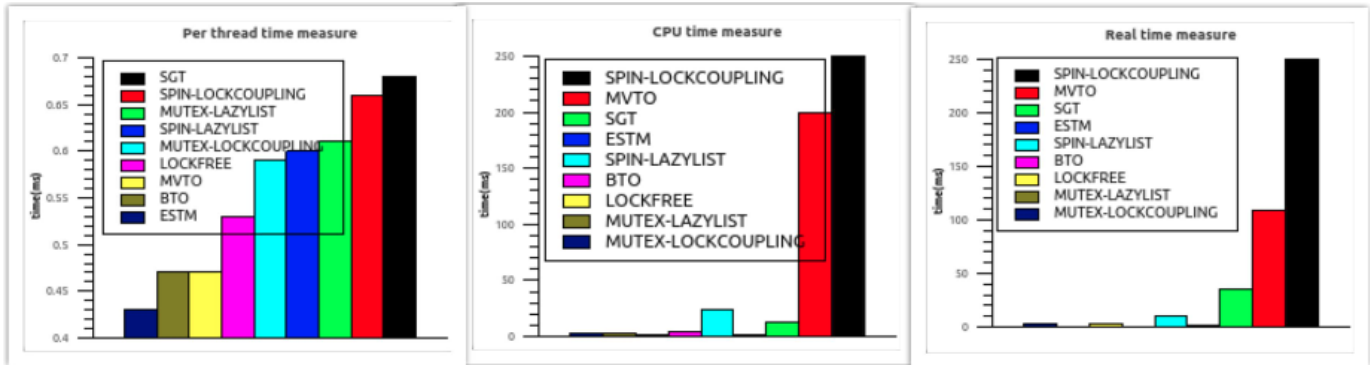

Fig. 9. All protocols comparison with different clock measures. Note number of threads is 100. Y axis is time(milliseconds)

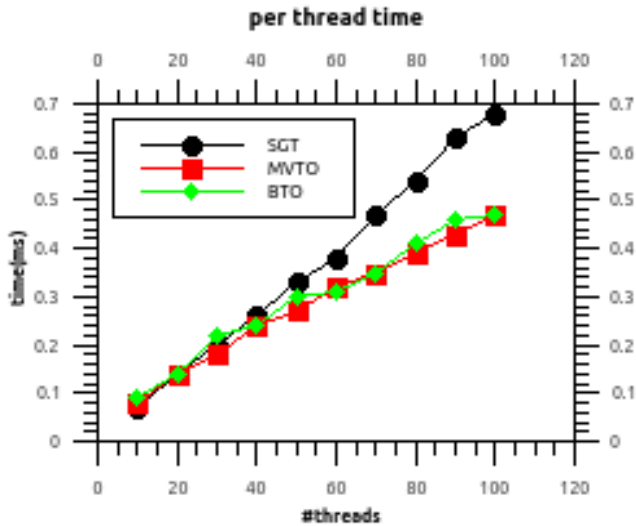

Fig. 10. SGT vs MVTO vs BTO. X-axis is the number of threads. Y axis is time taken in milliseconds

## V. RELATED WORK

A lot of STM systems exist in literature each one focusing on different design aspects, namely granularity of transactions, contention management, conflict resolution, synchronization, garbage collection, efficient data structure for metadata to enhancing concurrency and throughput of STMs [7]. Also, efficient maintenance of local logs, maintenance of versions of data objects, efficient method to validate the transactions and progress condition of transactions are some other factors influencing design of STMs.

None of these aspects alone are enough to design an efficient STM. Thus, the key to designing an efficient STM lies within selectively deciding on all these aspects and finding a midway to balance them to achieve desired performance of STM [4, 5, 6 and 7].

ENNALS STM [1] proposes an STM which aims to minimize cache contention by inline storage of object information. Here log of each transaction is local to it and is reclaimed as soon as a transaction commits. Writes take effect with 2 phase locking mechanism and reads are done optimistically.

Harris et al. [2] addresses the problem of bookkeeping overheads by avoiding local logging for lookups as their solution enables direct access of heaps rather than having local logs. It Introduces compile time optimizations. It also introduced for the first time the concept of storing version information alongside the objects rather than having separate version table. It uses garbage collection to reclaim the memory of obsolete objects.

DATM [3] proposes an efficient method of tracking dependencies of transactions, which enhances concurrency by accepting more transactions. It allows for the safe commit of transactions. TL2 [5] STM proposes lazy locking approach i.e. locking shared objects at commit time and combines it with a validation mechanism based on global version clock. TinySTM [6] uses eager locking, and to access shared memory it deploys an array of locks. A clock is implemented by shared counter. Besides this there has been a significant work to extend the classic software transactional model in terms of ESTM [8], ANT [11], open nested transactions [10] and Transactional boosting [9].

We use deferred write (lazy write) approach and implement STM by adapting state of art database concurrency protocols, namely SGT, time stamping and multi-version time-stamping protocols [13]. Synchrobench [12] contains an implementation of concurrent data structures via various concurrency control techniques. We mainly focus on *SET* implementation part of it. Interested readers can read Chen Fu et al. [7] where they beautifully give a detailed survey of STM systems based on transactional granularity, data organization, conflict detection, version management and synchronization.

## VI. Conclusion and Future work

We implemented an STM middleware with different concurrency control protocols (BTO, SGT and MVTO) and each of the protocol was tested against the different types of concurrency control protocols for the linked-list module of Synchrobench benchmark (lock coupling, lazy-list, lock-free list and ESTM) [12]. Synchrobench's linked-list module implements set using linked list. To test and compare performance we developed a test application using our STM middleware. For exhaustive evaluation, we used three clock measures: per thread CPU time, total CPU time and real time taken by the application.

Our experiments show that BTO performs better than MVTO which in turn is better than SGT for per thread CPU clock measure. But for total CPU time and real time clock measures SGT and BTO outperform MVTO. BTO and MVTO perform much better than lazy list and lock coupling set implementation of Synchrobench, when clock measure is per thread CPU time. ESTM is better than all other protocols across all clock measures.

We would like to extend this work by implementing another state of art concurrency control protocols and extend the STM middleware. We focused mainly on Set application in current work. We plan to do more experiments with other popular benchmarks (e.g. STAMP) and across different types of data structures (e.g. Tree, hash tables and skip lists). The source code of the current STM can be found at our research lab website [19].

## Acknowledgment

Our thanks to ANURAG labs, DRDO, department of defence, India for supporting the work. We would like to extend our sincere gratitude to anonymous reviewers for constructive suggestions.